\documentclass[aps, pra, showpacs, twocolumn]{revtex4}
\usepackage{graphics}
\usepackage{graphicx}
\usepackage{subfig}
\usepackage{amssymb}
\usepackage{amsmath}
\usepackage{bm}

\newcommand{\bra}[1]{\langle #1 \vert}
\newcommand{\ket}[1]{\vert #1 \rangle}

\begin{document}

\title{Quantum state reconstruction with imperfect rotations on an inhomogeneously broadened ensemble of qubits}
\author{Karl Tordrup}
\author{Klaus M{\o}lmer}
\affiliation{Lundbeck Foundation Theoretical Center for Quantum System Research, Department of Physics and Astronomy, University of  Aarhus,
DK-8000 Aarhus C, Denmark}

\date{\today}

\begin{abstract}
We present a method for performing quantum state reconstruction on qubits and qubit registers in the presence of decoherence and inhomogeneous
broadening. The method assumes only rudimentary single qubit rotations as well as knowledge of decoherence and loss mechanisms. We show that
full state reconstruction is possible even in the case where single qubit rotations may only be performed imperfectly. Furthermore we show that
for ensemble quantum computing proposals, quantum state reconstruction is possible even if the ensemble experiences inhomogeneous broadening and
if only imperfect qubit manipulations are available during state preparation and reconstruction.
\end{abstract}

\pacs{03.65.Wj, 03.67.Lx}

\maketitle

\section{Introduction}
The task of reconstructing a physical state from a set of measurement data is central to all branches of physics. Since full knowledge of the
quantum state allows only the prediction of a statistical distribution of outcomes one must perform numerous measurements on identically
prepared quantum systems in order to reconstruct the state. The abstract notion of the quantum mechanical  ensemble thus plays a very real part
in the business of quantum state reconstruction.

In this paper we shall consider the case of quantum state reconstruction within the framework of quantum computing. Since the ultimate goal of
quantum computing is to coherently steer the evolution of a quantum system towards a specific state wich encodes the answer to some
computational problem, the final state should be the single state that encodes the solution with unit (or very nearly unit) probability.
Presently quantum state reconstruction is an invaluable diagnostic tool for determining the fidelity with which basic quantum gates may be
performed.

A major challenge in all quantum computing systems is that of controlling the interactions between qubits. This problem is often tackled by
introducing an auxiliary degree of freedom. Some examples are ion trap quantum computing  where qubits are coupled via collective vibrational
modes of the trapped ions \cite{iontrap} and quantum computing with rare earth ions doped in inorganic crystals where qubit--qubit interaction
is mediated by a change in static dipole moment when a particular ion is promoted to an electronically excited state \cite{Kroll:optCommun}.
Thus although qubit dynamics are strictly speaking restricted to the qubit Hilbert space, many systems employ a larger space for the dynamics
with the restriction that at the end of a gate operation the effective evolution is described within the qubit space. This means that the
population of the auxiliary level and all coherences between the auxiliary level and the qubit levels should be sufficiently small at the end of
a gate operation and it makes sense to restrict oneself to the qubit two--level subsystem when performing quantum state reconstruction.

In quantum computing a lot of effort is put into performing qubit rotations (or gate operations) with as little error as possible. Naturally the
same sources of errors are present during the reconstructive procedure, however it is inappropriate that these errors should detract from the
gate fidelity. We therefore propose that decoherence and loss mechanisms be included in the theoretical account for the reconstruction. A
further complication occurs in ensemble quantum computing systems where a qubit consists of an entire ensemble of two--level systems with almost
identical physical characteristics. Qubit manipulations in such schemes must be robust to the natural variation of system parameters
\cite{Wesenberg:robust, NMR:robust, Khaneja}, but will inevitably lead to small differences in the evolution felt by each member of the
ensemble. During readout of such systems one obtains in fact the ensemble average of the measured quantity and the reconstructive procedure
should take this averaging into account.

The paper is arranged as follows. In Sec.\ \ref{sec:qubit} we analyze the case of single qubit state reconstruction. In Sec.\ \ref{sec:qubits}
we extend the results of Sec.\ \ref{sec:qubit} to include two-qubit registers. In Sec.\ \ref{sec:REQC} we present an application of our method
to a specific physical quantum computing implementation. Sec.\ \ref{sec:correl} discusses the case with correlations between the unknown quantum
state and reconstructive process and Sec.\ \ref{sec:conclusion} concludes the paper.

\section{Single qubit state reconstruction}\label{sec:qubit}
A conceptually simple way of reconstructing the quantum state of a two--level system is to apply a set of rotations $\{D_i\}$ and to measure the
expectation value of suitable observables in the rotated states. From the measured data and knowledge of the rotations performed one can infer
the full density operator of the unrotated state. For instance, in \cite{Sellars} coherent emission of radiation from rare-earth ions doped in
crystals provides information about quantum coherences and is thus applied to reconstruct various quantum states of the ions. For two qubits the
same procedure may be employed bearing in mind that the two qubits must be rotated differently in order to distinguish contributions from the
singlet and triplet states. The case where both qubits are subject to the same rotation is explored by Home \emph{et al.} in Ref.\ \cite{Home}.
In this case full reconstruction of the quantum state is not possible and it is necessary to resort to a numerical search.

We begin by recalling that the density matrix of any two--level system may be expanded on the set $\{\sigma_i\}$
\begin{equation}
\rho = \tfrac{1}{2}\sum_{j=0}^3 c_j \sigma_j.
\end{equation}
with $\sigma_{1-3}$ the Pauli matrices and $\sigma_0$ the identity. The condition $Tr(\rho)=1$ is enforced by setting $c_0=1$ and the further
condition that $\rho$ be positive leaves three real parameters fulfilling
\begin{equation}
\sum_{i=1}^3 c_i^2 \leq 1.
\end{equation}

\subsection{Qubit rotations}
In order to proceed with the reconstruction we must now perform a set of rotations on identically prepared copies of the initial state. When
dealing with a single qubit  the generic Hamiltonian in the standard basis $\{\ket{0},\ket{1}\}$ is given by
\begin{equation}\label{eq:H}
H = \tfrac{1}{2}\bm{\sigma}\cdot\bm{\hat{n}}\Omega,
\end{equation}
where $\bm{\sigma}$ is the vector of Pauli matrices and $\bm{\hat{n}}$ is a unit vector. When acting for a time t this Hamitonian effects a
rotation of the Bloch vector through $\Omega t$ about the vector $\hat{n}$. For a perfect rotation of the Bloch vector through $\theta_i$ about
an equatorial axis $\hat{n}=(\cos(\varphi_i),\sin(\varphi_i),0)$, the rotation operator is given by
\begin{equation}\label{eq:rot}
D_i = D(\varphi_i, \theta_i) = \begin{pmatrix}
                                              \cos(\theta_i / 2) & -ie^{-i\varphi_i}\sin(\theta_i /2) \\
                                              -ie^{i\varphi_i}\sin(\theta_i /2) & \cos(\theta_i /2)
                                \end{pmatrix}.
\end{equation}
Due to the linearity of quantum mechanics it is sufficient to know how the Pauli matrices transform under these rotations.
\begin{subequations}\label{eq:rotBasis}
\begin{eqnarray}
D\sigma_1D^{\dag}& = &\{\cos^2(\theta/2) + \sin^2(\theta/2)\cos(2\varphi)\}\sigma_1 \nonumber \\
                 &   &+ \sin^2(\theta/2)\sin(2\varphi)\sigma_2 \nonumber \\
                 &   &-\sin(\theta)\sin(\varphi)\sigma_3
\end{eqnarray}
\begin{eqnarray}
D\sigma_2D^{\dag} & = &\sin^2(\theta/2)\sin(2\varphi)\sigma_1 \nonumber \\
                  &  &+\{\cos^2(\theta/2)-\sin^2(\theta/2)\cos(2\varphi)\}\sigma_2 \nonumber \\
                  &  &+\sin(\theta)\cos(\varphi)\sigma_3
\end{eqnarray}
\begin{eqnarray}
D\sigma_3D^{\dag} & = &\sin(\theta)\sin(\varphi)\sigma_1 -\sin(\theta)\cos(\varphi)\sigma_2 \nonumber \\
                  &  &\cos(\theta)\sigma_3.
\end{eqnarray}
\end{subequations}
We can then infer how any initial state transforms under the rotation $D_i$,
\begin{eqnarray}\label{eq:rhoR}
\rho_R = D_i \rho D_i^{\dag}  &=& D_i (\sum_{j=1}^3 c_j \sigma_j) D_i^{\dag} \nonumber       \\
                              &=& \sum_{j=1}^3 c_j (D_i \sigma_j D_i^{\dag}) \nonumber      \\
                              &=& \sum_{j=1}^3 c_j (\sum_{k=1}^3 b_{jk} \sigma_k) \nonumber \\
                              &=& \sum_{j=1}^3 \sum_{k=1}^3 c_j b_{jk} \sigma_k,
\end{eqnarray}
where the $b_{jk}$ follow from Eqs.\ (\ref{eq:rotBasis}).

We shall find it convenient to represent the state simply by
\begin{equation}\label{eq:vrho}
\bm{v}_{\rho} = \begin{pmatrix}
                c_1 \\
                c_2 \\
                c_3 \\
                \end{pmatrix}
\end{equation}
From Eq.\ (\ref{eq:rhoR}) we may construct the superoperator
\begin{equation}
B = \begin{pmatrix}
    b_{11} & b_{21} & b_{31}\\
    b_{12} & b_{22} & b_{32}\\
    b_{13} & b_{23} & b_{33}
    \end{pmatrix},
\end{equation}
such that the rotated state may be found as
\begin{equation}\label{eq:vrhoR}
\bm{v}_{\rho}^R = B\bm{v}_{\rho}.
\end{equation}
Let us consider measurements which reveal the population of the $\ket{0}$ and $\ket{1}$ levels of the rotated state, i.e.
\begin{subequations}
\begin{align}
P_0 &= \tfrac{1}{2} + c_3^R\\
P_1 &= 1 - P_0 = \tfrac{1}{2} - c_3^R,
\end{align}
\end{subequations}
or equivalently using Eqs.\ (\ref{eq:vrho})--(\ref{eq:vrhoR})
\begin{equation}\label{eq:matrixEqBs}
\begin{pmatrix} b_{13} &  b_{23} &  b_{33} \\
                -b_{13} &  -b_{23} &  -b_{33}
\end{pmatrix}
\begin{pmatrix}
c_1 \\
c_2 \\
c_3
\end{pmatrix} =
\begin{pmatrix}
P_0 - \tfrac{1}{2}\\
P_1 - \tfrac{1}{2}
\end{pmatrix}.
\end{equation}
For the special case of Eq.\ (\ref{eq:rot}) we obtain the explicit equations
\begin{equation}\label{eq:matrixEq}
\begin{pmatrix} -\sin\theta\cos\varphi &  \sin\theta\cos\varphi &  \cos\theta \\
                 \sin\theta\cos\varphi & -\sin\theta\cos\varphi & -\cos\theta
\end{pmatrix}
\begin{pmatrix}
c_1 \\
c_2 \\
c_3
\end{pmatrix} =
\begin{pmatrix}
P_0 - \tfrac{1}{2}\\
P_1 - \tfrac{1}{2}
\end{pmatrix}.
\end{equation}
It is evident that the two equations are linearly dependent and thus one must choose at least three sets of rotation angles
$(\theta_i,\varphi_i)$ in order to obtain a coefficient matrix of rank 3. The coefficients $c_i$ are then found by simply inverting the
coefficient matrix of the resulting set of equations.

\subsection{Relaxation effects}
We now consider the complications of decoherence and coupling to auxiliary degrees of freedom. The problem of quantum state reconstruction in
damped systems is well studied, for instance a very general treatment is given in \cite{damping}. Here we shall assume that the decoherence
mechanisms and the coupling to auxiliary degrees of freedom are well understood and may thus be modelled accurately when describing the full
system dynamics which may take place in a larger space than the qubit space. The procedure now follows the same lines as previously. We choose a
set of rotations $\{(\theta,\phi)_i\}$ and identify a set of parameters describing decoherence effects. Taking these effects into account we
simulate the effect of each rotation on each of the basis states
\begin{equation}\label{eq:sigmaR}
\sigma_j^R = \sum_{k=0}^3b_{jk} \sigma_k,
\end{equation}
by numerical integration of the Lindblad master equation
\begin{equation}
\dot{\rho} = i[\rho,H]+\mathcal{L}_{\text{relax}}(\rho).
\end{equation}
The function $\mathcal{L}_{\text{relax}}$ given by \cite{Lindblad}
\begin{equation}
\mathcal{L}_{\text{relax}}(\rho) = -\tfrac{1}{2}\sum_m\{C_m^{\dag}C_m\rho+\rho C_m^{\dag}C_m\} + \sum_m C_m \rho C_m^{\dag},
\end{equation}
takes the relevant relaxation mechanisms into account. For instance spontaneous transitions from an auxiliary state $\ket{\text{aux}}$ to the
qubit state $\ket{0}$ with decay rate $\Gamma$ is modelled by a term with
\begin{equation}
C_m = \sqrt{\Gamma}\ket{0}\bra{\text{aux}}.
\end{equation}

Notice in Eq.\ (\ref{eq:sigmaR}) that the sum now runs from $j=0$ since some population may be lost to the auxiliary state forcing us to relax
the normalisation constraint and describe the state as a four component vector. From the calculated $b_{jk}$ and the measured populations $P_i$
for a single set of rotation angles $(\theta,\varphi)$ the equation for the $c_i$ is given by
\begin{equation}\label{eq:matrixEqBs}
\begin{pmatrix} b_{00}+b_{03} & b_{10}+b_{13} &  b_{20}+b_{23} &  b_{30}+b_{33} \\
                b_{00}-b_{03} & b_{10}-b_{13} &  b_{20}-b_{23} &  b_{30}-b_{33}
\end{pmatrix}
\begin{pmatrix}
c_0 \\
c_1 \\
c_2 \\
c_3
\end{pmatrix} =
\begin{pmatrix}
P_0 \\
P_1
\end{pmatrix}.
\end{equation}
Since we are now forced to relax the constraint $Tr(\rho) = 1$ because population may be lost to other levels the two equations of Eq.\
(\ref{eq:matrixEqBs}) are in general linearly independent. The minimum set of angles needed to reconstruct such a state is two.

\subsection{Ensemble quantum systems}
As a final level of complication we now consider the problem of state reconstruction in the context of ensemble quantum computing. Explicitly we
assume that each member of the ensemble reacts slightly differently to the external controls and that these differences may be quantified by
some set of parameters $\delta$. For example $\delta$ could signify a variation in the level structure brought on by the microscopic environment
around each ensemble member or a variation in the external control field over the spatial extent of the ensemble. When performing the qubit
rotations each member of the ensemble experiences a slightly different evolution from that of its neighbours so that with probability
$P(\delta)$ an ensemble member undergoes the evolution
\begin{equation}\label{eq:rhoDelta}
\rho \rightarrow \rho^R(\delta) = D(\theta,\varphi,\delta) \rho D^{\dag}(\theta,\varphi,\delta).
\end{equation}
The final state is given by the ensemble average
\begin{equation}
\overline{\rho^R} = \int \rho^R(\delta)P(\delta)d\delta.
\end{equation}

\begin{figure}
\includegraphics[width=4cm]{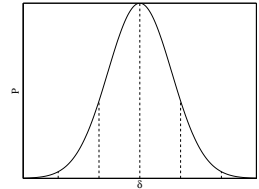}
\caption{The probability that a member of the ensemble experiences the error parameter value $\delta$ is given by
P($\delta$).}\label{fig:Pdistr}
\end{figure}

It is a valid point that the effects of $\delta$ would also be felt during the preparation of the state $\rho$. Means exist, however, to
eliminate the effects of error Hamiltonians that are constant in time and thus to compensate for variations in $\delta$ \cite{NMR:robust}. In
this section we shall thus assume that the ensemble represented by $\rho$ is uncorrelated with the value of $\delta$. In order to proceed with
the state reconstruction we choose a set of $N$ sample values of the variable $\delta$ as illustrated in Fig.\ \ref{fig:Pdistr}
and assign to each a weight $p_i$ proportional to the probability $P(\delta)$ such that
\begin{equation}
\sum_{i=1}^N p_i = 1.
\end{equation}
We then simulate the effects of a rotation $D(\theta,\varphi,\delta)$ on each basis state for each sample point
\begin{equation}
\sigma_j^R = \sum_{i=1}^N \sum_{k=0}^3 p_i b_{jk}\sigma_k.
\end{equation}
Again choosing two sets of angles $(\theta_i,\varphi_i)$ gives a linear system of equations analogous to Eq.\ (\ref{eq:matrixEqBs}) for the
$c_i$ which may be inverted to obtain the initial state $\bm{v}_{\rho}$.

\section{Two qubit state reconstruction}\label{sec:qubits}
In the previous section we asserted that a single qubit state can be expanded on the Pauli matrices. The state of a two qubit system may
therefore be expanded on the tensor product space
\begin{equation}
\rho = \sum_{i=0}^3 \sum_{j=0}^3 c_{ij} \sigma_i \otimes \sigma_j.
\end{equation}
The matrix representation of $\sigma_i \otimes \sigma_j$ is constructed as the Kronecker product of the two matrices $\sigma_i$ and $\sigma_j$.
The state vector $\bm{v}_{\rho}$ now becomes a 16 component vector of the $c_{ij}$. By applying rotations to both qubits we obtain the rotated
two--qubit state
\begin{eqnarray}
\rho_R  &=& (D_1 \otimes D_2) \rho (D_1^{\dag} \otimes D_2^{\dag}) \nonumber \\
        &=& (D_1 \otimes D_2) \sum_{i=0}^3 \sum_{j=0}^3 c_{ij} \sigma_i \otimes \sigma_j (D_1^{\dag} \otimes D_2^{\dag})\nonumber \\
        &=& \sum_{i=0}^3 \sum_{j=0}^3 c_{ij} (D_1 \sigma_i D_1^{\dag})\otimes(D_2 \sigma_j D_2^{\dag}) \nonumber \\
        &=& \sum_{i=0}^3 \sum_{j=0}^3 c_{ij} (\sum_{k=0}^3b_{ik}\sigma_k) \otimes (\sum_{k'=0}^3 b_{jk'}\sigma_{k'}) \nonumber \\
        &=& \sum_{i=0}^3 \sum_{j=0}^3 \sum_{k=0}^3 \sum_{k'=0}^3 c_{ij} b_{ik} b_{jk'} \sigma_k \otimes \sigma_{k'}.
\end{eqnarray}
As mentioned previously it is important that $D_1$ and $D_2$ implement different rotations in order to allow full reconstruction of the state.
As in the single qubit case we obtain an expression for the population of the basis state $\ket{nm}$
\begin{equation}\label{eq:Pnm}
P_{nm} = \sum_{i=0}^3 \sum_{j=0}^3 \sum_{k=\{0,3\}} \sum_{k'=\{0,3\}} sgn(n,m,k,k')c_{ij}b_{ik}b_{jk'},
\end{equation}
where $sgn(n,m,k,k')$ represents the sign of the entry $(n,m)$ in the tensor product $\sigma_k \otimes \sigma_{k'}$ with $k = 0,3$ and $k' =
0,3$. Eq.\ (\ref{eq:Pnm}) is analogous to Eq.\ (\ref{eq:matrixEqBs}) and the results of the previous section now extend naturally to the two
qubit case. We include the effects of decoherence and when working with an ensemble quantum computing system we include effects of inhomogeneity
over the ensemble in $\delta$ choosing again a set of sample points from the ensemble probability distribution as shown in Fig.\
\ref{fig:Pdistr}. For each set of angles $(\varphi_1,\theta_1,\varphi_2,\theta_2)$ we obtain four equations in the $c_{ij}$. In order to obtain
an invertible system of rank 16 we must thus choose at least four sets of angles $(\varphi_1,\theta_1,\varphi_2,\theta_2)$. This approach
assumes perfect measurement statistics which requires an infinite number of measurements in the laboratory and numerical precision on the order
of the machine precision during the integration of the master equation. In practice we choose at least five and preferably more sets of angles
and perform a least squares fit to the obtained data. A thorough description of quantum state reconstruction by least squares inversion is given
in \cite{damping}.

\section{Application to quantum computing with rare earth ions doped in crystals}\label{sec:REQC}
\begin{figure}
\includegraphics[width=8cm]{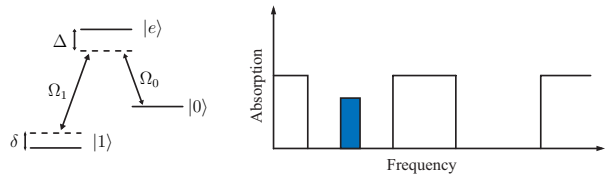}
\caption{Left: simplified level diagram of a single rare earth ion. Right: qubits are prepared as peaks in a hole burnt
structure.}\label{fig:lambdaSys}
\end{figure}

As an example we shall now apply the methods developed in the previous sections to the case of quantum state reconstruction in the context of
quantum computing with rare earth ions doped in inorganic crystals. We shall base our study on a rare earth quantum computing proposal first
developed by Ohlsson \emph{et al.} \cite{Kroll:optCommun}. In this proposal qubit states are encoded in hyperfine levels of the rare earth ions
embedded in the crystal. The qubit levels are coupled via an electronically excited state accessible via optical fields. Due to the large
inhomogeneous broadening of the excited state qubits may be defined by isolated peaks in frequency space as seen in Fig.\ \ref{fig:lambdaSys}.
Each qubit thus consists of an ensemble of ions centered at a given transition frequency. Qubit--qubit interaction is mediated by the
dipole--dipole coupling between different ions.
In rare earth quantum computing arbitrary single qubit rotations as described by Eq.\ (\ref{eq:rot}) may be implemented using two fields
$\Omega_0$ and $\Omega_1$ simultaneously \cite{Roos}. The implementation of single qubit rotations is made robust to variations in the excited
state shift $\Delta$, within the width of the qubit channel (see Fig.\ \ref{fig:lambdaSys}), by sweeping the frequency of the laser across the
width of the qubit in frequency space. Typical parameters for the system are $\Omega_0 \sim \Omega_1 \sim 1$ MHz, $\Delta \sim 100$ kHz and
$\delta \sim 10$ kHz. Recently we have shown that errors due to the inhomogeneous broadening $\delta$ of the hyperfine splitting between the
qubit levels are in fact dynamically suppressed \cite{selfREQC} during single qubit rotations, but they cannot be made vanishingly small.

\begin{figure}
\includegraphics[width=8cm]{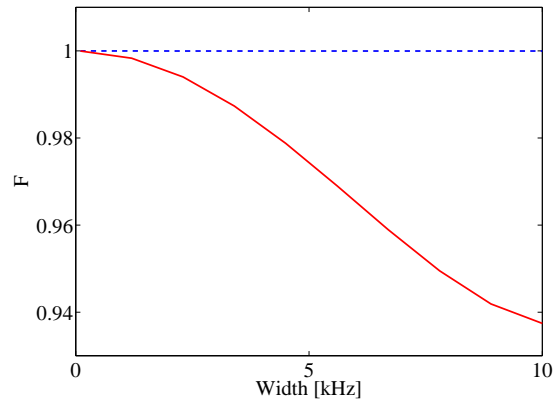}
\caption{(Color online) Fidelity of the quantum state reconstruction of a two qubit state as a function of the width of the inhomogeneous
profile. The dashed line is found by averaging correctly over the ensemble while the full line is found by ignoring variations over the
ensemble.}\label{fig:F2}
\end{figure}

In the following we simulate reconstruction of the qubit state $\ket{\psi}=1/\sqrt{2}(\ket{0}+\ket{1})$ using three sets of rotation angles. The
inhomogeneous profile is modelled as a Lorentzian with variable width and we evaluate the fidelity of the reconstruction as
\begin{equation}
F = \bra{\psi}\tilde{\rho}\ket{\psi},
\end{equation}
where $\tilde{\rho}$ is the reconstructed state. In order to evaluate the usefulness of our method we first perform the reconstruction taking
the ensemble averaging into account and compare this with the reconstruction assuming $\delta=0$ for all ensemble members. In the latter case
the linear equation system of Eq.\ (\ref{eq:matrixEqBs}) is not exactly soluble so a least squares fit is performed.
\begin{figure}
\includegraphics[width=8cm]{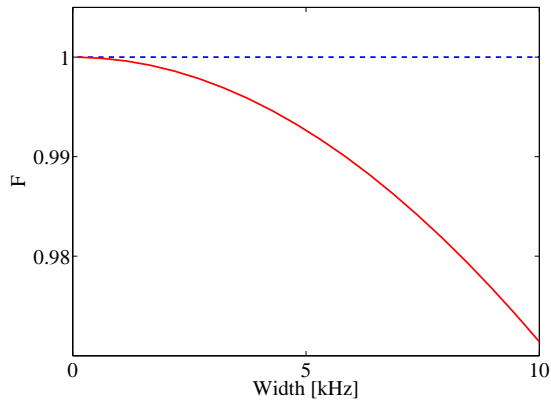}
\caption{(Color online) Fidelity of the quantum state reconstruction of a single qubit state as a function of the width of the inhomogeneous
profile. The dashed line is found by averaging correctly over the ensemble while the full line is found by ignoring variations over the
ensemble.}\label{fig:F1}
\end{figure}
In Fig.\ \ref{fig:F1} we plot the fidelity as a function of the width of the inhomogeneous profile. As expected the fidelity drops when the
width of the Lorentzian increases if we do not properly include the $\delta$ dependent contributions to the final state. We have also simulated
the reconstruction of the fully entangled two qubit state $\ket{\psi}=1/2(\ket{00}+\ket{01}+\ket{10}-\ket{11})$ using six sets of angles. In
Fig.\ \ref{fig:F2} we plot the fidelity as a function of the width of the inhomogeneous profile. In this case the drop in fidelity as the width
of the Lorentzian increases is significant if ensemble averaging is disregarded and failure to take this into account during the reconstructive
procedure may lead to overly pessimistic estimates of the fidelity of the operation producing the entangled state.

\section{Correlated perturbations in preparation and reconstruction stages}\label{sec:correl}

\begin{figure} \centering
\subfloat[] {\includegraphics[width=4cm]{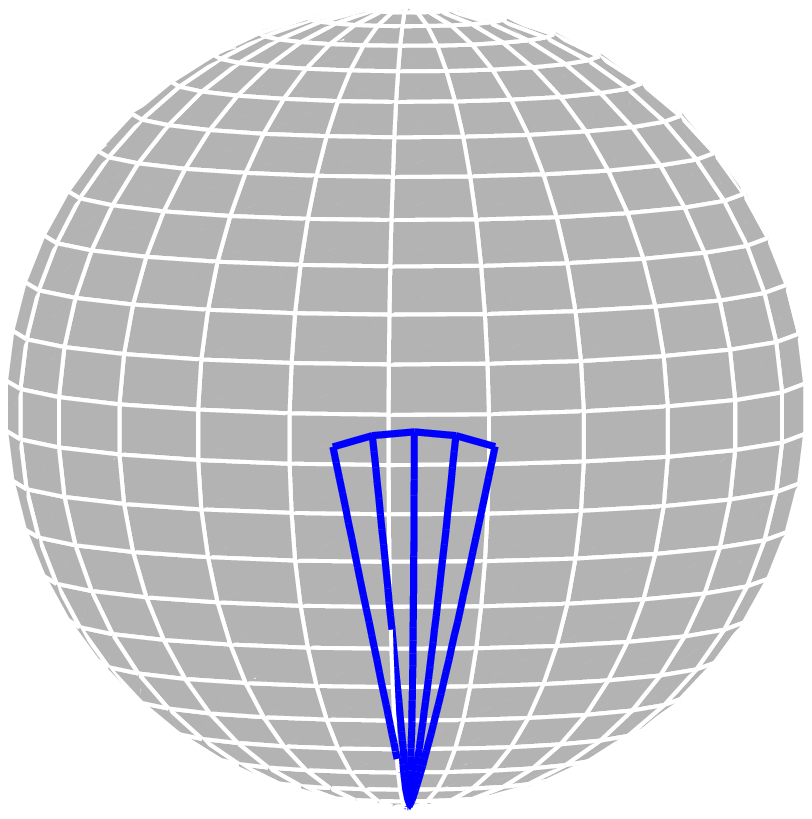}\label{fig:Bloch1}}  \subfloat[] {\includegraphics[width=4cm]{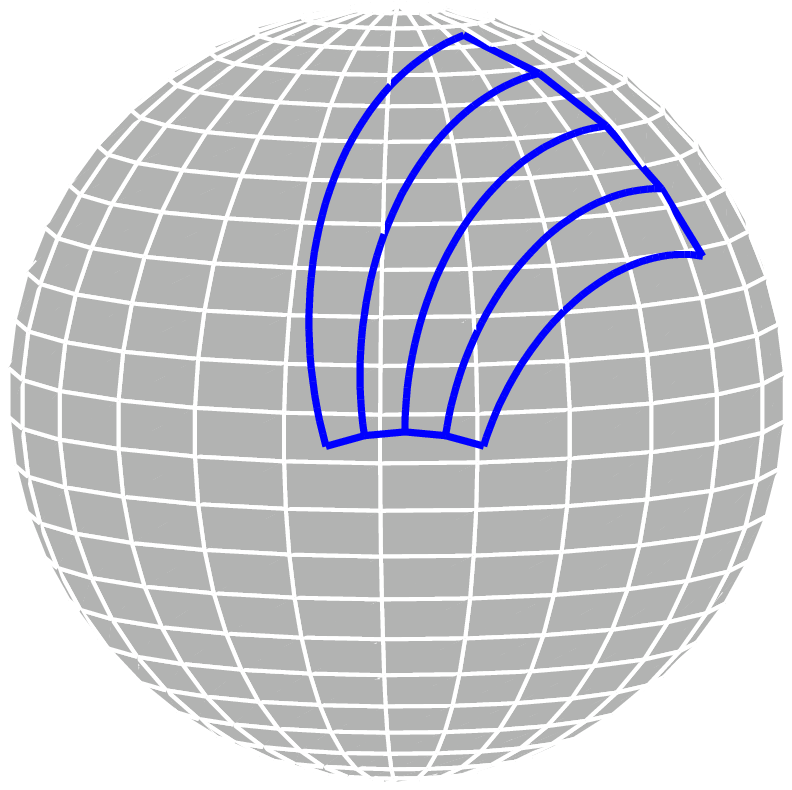}\label{fig:Bloch2}}
\caption{(Color online) (a): Applying a gate operation to a pure state qubit causes ensemble members to diverge on the Bloch sphere. Only a few
representative paths corresponding to the dashed lines in Fig. \ref{fig:Pdistr} are shown. (b): In the subsequent reconstructive rotations the
paths followed by different ensemble members are correlated with the paths followed during the gate rotation.}\label{fig:Blochs}
\end{figure}

Consider a single qubit ensemble in the pure spin down state $\bm{v}_{\rho}^{(i)} = (0,0,-1)$. Performing a gate operation corresponds to
rotating the Bloch vector as illustrated in Fig.\ \ref{fig:Bloch1}. Each member of the ensemble follows a distinct path along the Bloch sphere
and the final state $\bm{v}_{\rho}$ is the mixed state that results from taking the statistical average of all states along the upper blue line.
If we now wish to reconstruct $\bm{v}_{\rho}$ we must perform a number of rotations as exemplified in Fig.\ \ref{fig:Bloch2}, but since each
ensemble member experiences the same error during the tomographic rotation as during the initial rotation, the path followed on the Bloch sphere
by each member is correlated to the path followed during the initial rotation.
It is important to verify that these correlations do not preclude the identification of the state $\bm{v}_{\rho}$. Intuitively it should be
clear from Fig.\ \ref{fig:Blochs} that a similar problem occurs no matter which aspect of the Hamiltonian of Eq.\ (\ref{eq:H}) is affected by
$\delta$. Any aspect of the external control that varies with $\delta$ will cause different ensemble members to follow different paths.

In the preceding sections we assumed that the state to be reconstructed did not posses any correlations between the quantum state and $\delta$.
Whether dealing with an ensemble quantum computing system where a single qubit is encoded in an ensemble of physical two-level systems or
"single instance" systems where a single qubit is encoded in a single physical two--level system, this case is well described by the quantum
operations formalism \cite{operatorSum}
\begin{equation}
\rho \rightarrow \sum_k E_k \rho E_k^{\dag}.
\end{equation}
The $E_k$ are linear operators on the qubit Hilbert space fulfilling the relation
\begin{equation}
\sum_k E_k E_k^{\dag} = 1.
\end{equation}
In the ensemble case studied above where ensemble members undergo with probability $P(\delta)$ the unitary evolution $U_{\delta}$ we have simply
\begin{equation}
E_k = \sqrt{P(\delta)}U_{\delta}.
\end{equation}
This is the case depicted graphically in Fig.\ \ref{fig:Bloch1}. We shall now consider the case depicted in Fig.\ \ref{fig:Bloch2} where the
evolution of each ensemble member is correlated to the distinct evolution experienced by that member during the preparation of  the state. This
case is \emph{not} contained within the quantum operations formalism and so it is not \emph{a priori} clear that state reconstruction is
possible in this case. We shall now prove that full state reconstruction is in fact possible. We can assume that at some time in the past the
ensemble was in a state possessing no correlations with $\delta$. The cumulative effects of all subsequent manipulations including the gate
operation used to prepare the state is that with probability $P(\delta)$ a given ensemble member evolves into $\bm{v}_{\rho}(\delta)$. We
proceed by expanding $\bm{v}_{\rho}(\delta)$ in powers of the inhomogeneity parameter $\delta$
\begin{equation}\label{eq:initExp}
\bm{v}_{\rho}(\delta) = \bm{v}_{\rho}^{(0)} + \delta \bm{v}_{\rho}^{(1)} + \delta ^2 \bm{v}_{\rho}^{(2)} + \mathcal{O}(\delta^3).
\end{equation}
We also expand the superoperator B describing the tomographic rotation in powers of $\delta$
\begin{equation}
B(\delta) = B^{(0)} + \delta B^{(1)} + \delta^2 B^{(2)} + \mathcal{O}(\delta^3).
\end{equation}
We may think of each ensemble member as being finally rotated with probability $P(\delta)$ into the state
\begin{eqnarray}
\bm{v}_{\rho}^R(\delta) &=& (B^{(0)} + \delta B^{(1)} + \delta^2 B^{(2)})(\bm{v}_{\rho}^{(0)} + \delta \bm{v}_{\rho}^{(1)} + \delta^2 \bm{v}_{\rho}^{(2)})\nonumber\\
       &=& B^{(0)}\bm{v}_{\rho}^{(0)} + \delta (B^{(0)}\bm{v}_{\rho}^{(1)} + B^{(1)}\bm{v}_{\rho}^{(0)}) \nonumber\\
       & &+ \delta^2(B^{(1)}\bm{v}_{\rho}^{(1)} + B^{(2)}\bm{v}_{\rho}^{(0)} + B^{(0)}\bm{v}_{\rho}^{(2)}),
\end{eqnarray}
where we have kept only terms up to $\mathcal{O}(\delta^2)$. When performing a measurement on the rotated state we therefore obtain the ensemble
average
\begin{eqnarray}\label{eq:finalExp}
\overline{\bm{v}_{\rho}^R} &=& \int P(\delta)\bm{v}_{\rho}^R(\delta)d\delta \nonumber \\
                  &=& B^{(0)}\bm{v}_{\rho}^{(0)} + \overline{\delta} (B^{(0)}\bm{\rho}^{(1)} + B^{(1)}\bm{v}_{\rho}^{(0)}) \nonumber\\
       & &+ \overline{\delta^2}(B^{(1)}\bm{v}_{\rho}^{(1)} + B^{(2)}\bm{v}_{\rho}^{(0)} + B^{(0)}\bm{v}_{\rho}^{(2)}).
\end{eqnarray}
If the distribution $P(\delta)$ is symmetric around the origin the first order term averages to zero. However from Eq.\ (\ref{eq:finalExp}) it
is evident that $\bm{v}_{\rho}$ may be reconstructed to second order simply by expanding the rotation operator to the same order. Similarly the
state can in fact be found to any order as long as the rotation operator is known to the same order. In order to determine the individual terms
in the expansion of the state given in Eq.\ (\ref{eq:initExp}) one must employ successively more sets of rotation angles. If, however we are
only interested in the measured ensemble average the discussion of Sec.\ \ref{sec:qubits} still holds. In this case the expansion of the state
is contained implicitly in the averaging over $\delta$ and four sets of angles is the minimum to reconstruct the state. In a single instance
quantum computing proposal the above arguments show that quantum state reconstruction can be performed even in systems with intrinsic systematic
errors quantified by $\delta$.

\section{conclusion}\label{sec:conclusion}
In conclusion we have described a simple and effective method of performing quantum state reconstruction in the context of imperfect quantum
computing. The method can be adapted to include decoherence effects inherent in a given specific quantum computing proposal. Moreover we have
shown that full quantum state reconstruction is possible even in systems that interact with environmental degrees of freedom that were also
important in the preparation of the state. This is a good example of a process which is not contained within the quantum operations formalism,
and we note that such a scenario was already mentioned with a rather contrived yet illuminating example in \cite{Nielsen}.

\bibliography{minbib}

\end{document}